\let\ifarxiv=\iftrue     % ARXIV VERSION
\pdfoutput=1

\ifarxiv

\documentclass[12pt,a4paper]{article}
\usepackage[a4paper,text={450pt,650pt},centering]{geometry}

\fi

\ifarxiv\else

\documentclass[11pt,a4paper]{article}
\usepackage{mathptmx}
\usepackage[a4paper,text={130mm,198mm}]{geometry}

\fi

\usepackage{bbm} 
\usepackage{slashed} 
\usepackage{color}
\newcommand{\gray}[1]{\textcolor[gray]{0.5}{#1}}

\usepackage{amsmath,amssymb}
\usepackage[bookmarks=true,hyperfigures=true]{hyperref}
\usepackage{graphicx}
\usepackage[nosort]{cite}
\usepackage[bulletsep]{collref}
\let\oldbfseries=\bfseries
\let\oldmdseries=\mdseries
\let\oldnormalfont=\normalfont
\renewcommand{\bfseries}{\oldbfseries\boldmath}
\renewcommand{\mdseries}{\oldmdseries\unboldmath}
\renewcommand{\normalfont}{\oldnormalfont\unboldmath}

\allowdisplaybreaks[3]

\numberwithin{equation}{section}

\usepackage[font=small,labelfont=bf,width=0.85\textwidth]{caption}

\providecommand{\hypersetup}[1]{}
\providecommand{\texorpdfstring}[2]{#1}

\hypersetup{plainpages=false}
\hypersetup{pdfpagemode=UseNone}
\hypersetup{bookmarksnumbered=true}
\hypersetup{pdfstartview=FitH}
\hypersetup{colorlinks=false}
\hypersetup{citebordercolor={.5 1 .5}}
\hypersetup{urlbordercolor={.5 1 1}}
\hypersetup{linkbordercolor={1 .7 .7}}
\providecommand{\href}[2]{#2}
\providecommand{\arxivlink}[1]{\href{http://arxiv.org/abs/#1}{arxiv:#1}}

\newcommand{\superN}{\mathcal{N}}
\newcommand{\Action}{\mathcal{S}}

\newcommand{\tr}{\mathop{\mathrm{tr}}}

\newcommand{\str}{\mathop{\mathrm{str}}}

\newcommand{\Integers}{\mathbbm{Z}}
\newcommand{\Reals}{\mathbbm{R}}
\newcommand{\Complex}{\mathbbm{C}}
\newcommand{\Sphere}{S}  % {\mathbbm{S}}
\newcommand{\AdS}{\mathrm{AdS}}
\newcommand{\CP}{\mathrm{CP}}
\newcommand{\RP}{\mathrm{RP}}

\ifx\genfrac\sdflkaj

\else

\fi
\newcommand{\sfrac}[2]{{\textstyle\frac{#1}{#2}}}
\newcommand{\half}{\sfrac{1}{2}}

\newcommand{\quarter}{\sfrac{1}{4}}

\newcommand{\rep}[1]{{\mathbf{#1}}}

\newcommand{\grp}[1]{\mathrm{#1}}

\newcommand{\grU}{\grp{U}}
\newcommand{\grSU}{\grp{SU}}
\newcommand{\grSO}{\grp{SO}}
\newcommand{\grSp}{\grp{Sp}}
\newcommand{\grSL}{\grp{SL}}
\newcommand{\grPSU}{\grp{PSU}}
\newcommand{\grOSp}{\grp{OSp}}

\newcommand{\lrbrk}[1]{\left(#1\right)}
\newcommand{\bigbrk}[1]{\bigl(#1\bigr)}

\newcommand{\biggbrk}[1]{\biggl(#1\biggr)}

\newcommand{\Bigsbrk}[1]{\Bigl[#1\Bigr]}

\newcommand{\Biggsbrk}[1]{\Biggl[#1\Biggr]}

\newcommand{\abs}[1]{{|#1|}}

\newcommand{\comma}{\quad,\quad} 
\newcommand{\levi}{\epsilon}

\newcommand{\nn}{\nonumber}

\newcommand{\nl}[1][0pt]{\nonumber\\[#1]&\hspace{-4\arraycolsep}&\mathord{}}

\newcommand{\earel}[1]{\mathrel{}&\hspace{-2\arraycolsep}#1\hspace{-2\arraycolsep}&\mathrel{}}
\newcommand{\eq}{\earel{=}} 
\newcommand{\be}{\begin{eqnarray}}
\newcommand{\ee}{\end{eqnarray}}

\makeatletter
\def\mr@ignsp#1 {\ifx\:#1\@empty\else #1\expandafter\mr@ignsp\fi}%
\newcommand{\multiref}[1]{\begingroup%\let\protect\string%
\xdef\mr@no@sparg{\expandafter\mr@ignsp#1 \: }%
\def\mr@comma{}%
\@for\mr@refs:=\mr@no@sparg\do{\mr@comma\def\mr@comma{,}\ref{\mr@refs}}%
\endgroup}
\makeatother

\newcommand{\secref}[1]{Sec.~\multiref{#1}}

\newcommand{\tabref}[1]{Tab.~\multiref{#1}}

\newcommand{\figref}[1]{Fig.~\multiref{#1}}
\begin{document}

%%%%%%%%%%%%%%%%%%%%%%%%%%%%%%%%%%%%%%%%%%%%%%%%%%%%%%%%%%%%%%%%%%%%%%%%%%%%%%%%
%%%%%%%%%%%%%%%%%%%%%%%%%%%%%%%%%%%%%%%%%%%%%%%%%%%%%%%%%%%%%%%%%%%%%%%%%%%%%%%%
% TITLE PAGE

\thispagestyle{empty}
\phantomsection
\addcontentsline{toc}{section}{Title}

\begin{flushright}\footnotesize%
\texttt{UUITP-37/10},
\texttt{\arxivlink{1012.3999}}\\
overview article: \texttt{\arxivlink{1012.3982}}%
\vspace{1em}%
\end{flushright}

\begingroup\parindent0pt
\begingroup\bfseries\ifarxiv\Large\else\LARGE\fi
\hypersetup{pdftitle={Review of AdS/CFT Integrability, Chapter IV.3: N=6 Chern-Simons and Strings on AdS4xCP3}}%
Review of AdS/CFT Integrability, Chapter IV.3:\\
$\mathcal{N}$ = 6 Chern-Simons and Strings on $AdS_4 \times CP^3$
\par\endgroup
\vspace{1.5em}
\begingroup\ifarxiv\scshape\else\large\fi%
\hypersetup{pdfauthor={Thomas Klose}}%
Thomas Klose
\par\endgroup
\vspace{1em}
\begingroup\itshape
Department of Physics and Astronomy, Uppsala University \\
SE-751 08 Uppsala, Sweden 
\par\endgroup
\vspace{1em}
\begingroup\ttfamily
thomas.klose@physics.uu.se 
\par\endgroup
\vspace{1.0em}
\endgroup

\begin{center}
\includegraphics[width=5cm]{TitleIV3.mps}%figure for your chapter
\vspace{1.0em}
\end{center}

\paragraph{Abstract:}
We review the duality and integrability of $\superN = 6$ superconformal Chern\discretionary{-}{-}{-}Simons theory in three dimensions and IIA superstring theory on the background $\AdS_4\times\CP^3$. We introduce both of these models and describe how their degrees of freedom are mapped to excitations of a long-range integrable spin-chain. Finally, we discuss the properties of the Bethe equations, the S-matrix and the algebraic curve that are special to this correspondence and differ from the case of $\superN=4$ SYM theory and strings on $\AdS_5\times\Sphere^5$. 

\ifarxiv\else
% http://www.ams.org/msc
\fi
\hypersetup{pdfsubject={MSC (2010): 81T13, 81T30, 37K10}}%

\ifarxiv\else
\fi
\hypersetup{pdfkeywords={Superconformal Chern-Simons theory, superstrings on AdS4xCP3, AdS/CFT correspondence, integrability}}%

\newpage

%%%%%%%%%%%%%%%%%%%%%%%%%%%%%%%%%%%%%%%%%%%%%%%%%%%%%%%%%%%%%%%%%%%%%%%%%%%%%%%%
%%%%%%%%%%%%%%%%%%%%%%%%%%%%%%%%%%%%%%%%%%%%%%%%%%%%%%%%%%%%%%%%%%%%%%%%%%%%%%%%
% BODY

%%%%%%%%%%%%%%%%%%%%%%%%%%%%%%%%%%%%%%%%%%%%%%%%%%%%%%%%%%%%%%%%%%%%%%%%%%%%%%%%
\section{Introduction}

Almost all statements that have been made in the other chapters of this review \cite{chapIntro} about the duality and integrability of string theory on $\AdS_5\times\Sphere^5$ and $\superN=4$ Yang-Mills theory in four dimensions, also hold in an appropriately adopted form for a second example of the AdS/CFT correspondence. This example has been known since June 2008 \cite{Aharony:2008ug}, and it is as concrete as the ``old'' one. Because the involved space-times are of one less dimension, this correspondence is often referred to as AdS$_4$/CFT$_3$ to distinguish it from the more established AdS$_5$/CFT$_4$.\footnote{Since December 2009, also an AdS$_3$/CFT$_2$ correspondence has been discussed in the context of integrability \cite{Babichenko:2009dk}.}

In the AdS$_5$/CFT$_4$ case, we had IIB superstring theory on $\AdS_5\times\Sphere^5$ with self-dual RR 5-form flux $F^{(5)} \sim N$ through $\AdS_5$ and $\Sphere^5$. This is now replaced by:
\be \label{eqn:IIA-theory}
\parbox{100mm}{
IIA superstring theory on $\AdS_4\times\CP^3$\\
with RR four-form flux $F^{(4)} \sim N $ through $\AdS_4$\\
and RR two-form flux $F^{(2)} \sim k$ through a $\CP^1\subset\CP^3$.
}
\ee
On the gauge theory side, we had $\superN=4$ superconformal Yang-Mills theory with coupling $g_{\mathrm{YM}}$ and gauge group $\grU(N)$ on $\Reals^{1,3}$. Now this is replaced by ABJM theory:
\be \label{eqn:ABJM-theory}
\parbox{100mm}{
$\superN=6$ superconformal Chern-Simons-matter theory\\
with gauge group $\grU(N)\times\grU(N)$ on $\Reals^{1,2}$\\
and Chern-Simons levels $k$ and $-k$.
}
\ee

Both theories are controlled by two and only two parameters, $k$ and $N$, which take integer values. These parameters determine all other quantities like coupling constants and the effective string tension. In ABJM theory, the Chern-Simons level $k$ acts like a coupling constant. The fields can be rescaled in such a way that all interactions are suppressed by powers of $\frac{1}{k}$, i.e.\ large $k$ is the weak coupling regime. 
%We will see this explicitly when we discuss the ABJM action in section \ref{sec:ABJM-theory}.
One can take a planar, or 't Hooft, limit which is given by
\be \label{eqn:tHooft}
  k,N \to \infty \comma \lambda \equiv \frac{N}{k} = \mathrm{fixed} \; .
\ee
It is in this limit where integrability shows up and which is therefore the focus of this review. On the gravity side, the string coupling constant and effective tension are given by\footnote{There are corrections to the second relation at two loops in the sigma model \cite{Bergman:2009zh}.}
\be \label{eqn:stringparameter}
  g_s \sim \lrbrk{\frac{N}{k^5}}^{1/4} = \frac{\lambda^{5/4}}{N}
  \comma
  \frac{R^2}{\alpha'} = 4\pi\sqrt{2\lambda}
  \; ,
\ee
where $R$ is the radius of $\CP^3$ and \emph{twice} the radius of $\AdS_4$. These relations are qualitatively the same as in the AdS$_5$/CFT$_4$ context. In the planar limit $g_s$ goes to zero and thus the strings do not split or join. At small 't Hooft coupling, the background is highly curved and the string is subject to large quantum fluctuations. At large 't Hooft coupling, the background is weakly curved which renders the sigma-model weakly coupled and the string behaves classically.

The first equation in \eqref{eqn:stringparameter} contains a hint that the duality is about more than the relationship between \eqref{eqn:IIA-theory} and \eqref{eqn:ABJM-theory}. If we are not in the 't Hooft limit but if we let $N \gg k^5$, then the string coupling $g_s$ becomes large. However, strongly coupled IIA string theory is M-theory. Indeed, ABJM theory \eqref{eqn:ABJM-theory} at arbitrary value of $k$ and $N$ is dual to \cite{Aharony:2008ug}
\be \label{eqn:M-theory}
\parbox{85mm}{
M-theory on $\AdS_4 \times \Sphere^7/\Integers_k$\\
with four-form flux $F^{(4)} \sim N $ through $\AdS_4$.
}
\ee
In other words, ABJM theory is the world-volume theory of a stack of $N$ M2 branes moving on $\Complex^4/\Integers_k$ \cite{Aharony:2008ug}. The duality of \eqref{eqn:IIA-theory} and \eqref{eqn:ABJM-theory} is really only a corollary of this more general M/ABJM duality in the limit where $k^5 \gg N$ and where therefore M-theory is well approximated by weakly coupled IIA string theory on a $\AdS_4 \times \CP^3$ background\footnote{$CP^3$ arises from writing $\Sphere^7$ as $\Sphere^1$ fibered over $\CP^3$ and by identifying the circle as the M-theory direction which shrinks to zero size by the orbifold action of $\Integers_k$ in the large $k$ limit.}. The lecture notes \cite{Klebanov:2009sg} discuss the general M/ABJM correspondence. However, in the planar limit \eqref{eqn:tHooft}, where $k$ and $N$ grow large with equal powers, we are always in the IIA regime. Thus, by concentrating on the question of integrability we are only concerned with IIA/ABJM. An extended and largely self-contained review of the AdS$_4$/CFT$_3$ correspondence is forthcoming \cite{newreview}.

\paragraph{Overview.} In a nutshell, the differences between AdS$_5$/CFT$_4$ and AdS$_4$/CFT$_3$, see \tabref{tab:nutshell}, are: The first duality involves theories that are invariant under the supergroup $\grPSU(2,2|4)$ and therefore are maximally supersymmetric (32 supercharges), while the theories in the second duality are $\grOSp(6|4)$-symmetric, a group which contains ``only'' 24 supercharges. After gauge fixing, the symmetry groups reduce to two and one copy of $\grSU(2|2)$, respectively. The \emph{sixteen} elementary excitations in the 5/4d case transform in the representation $(2|2)_L\otimes(2|2)_R$ of the residual symmetry group, while there are only \emph{eight} elementary excitations in the 4/3d case which transform in the representation 
\be \label{eqn:A+B}
  (2|2)_{A\mathrm{-particles}}\oplus(2|2)_{B\mathrm{-particles}} \; .
\ee
In \secref{sec:ABJMtoInt} and \secref{sec:IIAtoInt} we will show how these two types of particles arise from the gauge and string theory degrees of freedom, respectively.

\begin{table}%
\begin{center}
\begin{tabular}{l|c|c}
                  & AdS$_5$/CFT$_4$                                  & AdS$_4$/CFT$_3$ \\[1mm] \hline
                  && \\[-3mm]
Global symmetry   & $\grPSU(2,2|4)$                                  & $\grOSp(6|4)$ \\[2mm]
\raisebox{11mm}[0mm]{Dynkin diagram}
                  & \raisebox{6mm}[0mm]{\includegraphics[scale=0.4]{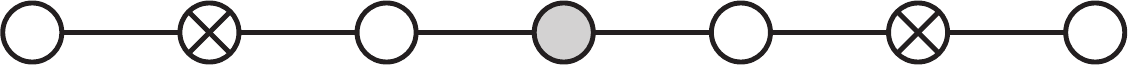}}
                                                                     & \includegraphics[scale=0.4]{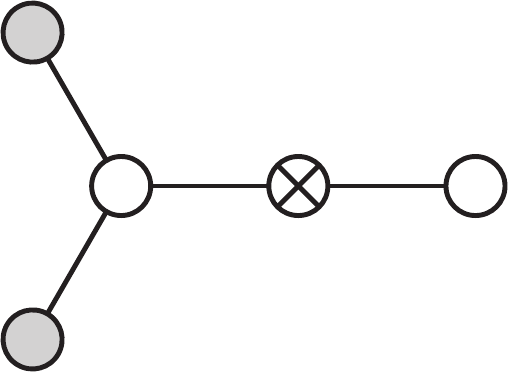} \\[3mm]
Residual symmetry & $\grSU(2|2)_L\times\grSU(2|2)_R$                 & $\grSU(2|2)$ \\[3mm]
Representations   & $(2|2)_L\otimes(2|2)_R = 16\, \mathrm{d.o.f}$    & $(2|2)_A\oplus(2|2)_B = 8\, \mathrm{d.o.f}$
\end{tabular}
\end{center}
\caption{\textbf{Comparison of symmetries.} The Dynkin diagram of $\grPSU(2,2|4)$ contains two $\grSU(2|2)$ branches which represent the residual symmetries, and exactly one momentum carrying root which we marked by shading it gray. This indicates that all 16 elementary excitations transform in a single irreducible representation with one fundamental index in each $\grSU(2|2)$. The Dynkin diagram of $\grOSp(6|4)$ contains only one $\grSU(2|2)$ branch, but two momentum carrying roots. Correspondingly, the 8 elementary excitations transform in two copies of the fundamental representation of $\grSU(2|2)$.}
\label{tab:nutshell}
\end{table}

Another difference between the two dualities is that the interpolation between weak and strong coupling in AdS$_4$/CFT$_3$ is much more intricate. Take e.g.\ the magnon dispersion relation, which due to the underlying $\grSU(2|2)$ symmetry is fixed in either duality to be of the form \cite{Beisert:2005tm} (see also \cite{Berenstein:2008dc})
\be \label{eqn:general-disp-rel}
  E(p) = \sqrt{Q^2 + 4 h^2(\lambda) \sin^2\tfrac{p}{2}} \; ,
\ee
where $Q$ is the magnon R-charge and where the function $h(\lambda)$ is \emph{not} fixed by symmetry. The fundamental magnon in AdS$_5$/CFT$_4$ has charge $Q=1$, while in AdS$_4$/CFT$_3$ it has $Q=\half$. In the AdS$_5$/CFT$_4$ case the function $h(\lambda)$ turned out to be simply $\sqrt{\lambda}/4\pi$, which can be argued to arise from S-duality \cite{Berenstein:2009qd}. In the present case there is no such argument and indeed the function $h$ happens to be quite non-trivial. The weak and strong coupling asymptotics are given by
\be \label{eqn:general-h-expansion}
  h(\lambda) = \begin{cases} 
    \lambda \Bigsbrk{ 1 + c_1 \lambda^2 + c_2 \lambda^4 + \ldots }         & \mbox{for $\lambda\ll 1$} \; , \\[2mm]
    \sqrt{\frac{\lambda}{2}} + a_1 + \frac{a_2}{\sqrt{\lambda}} + \ldots   & \mbox{for $\lambda\gg 1$} \; ,
  \end{cases}
\ee
where the leading terms were deduced in \cite{Minahan:2008hf,Gaiotto:2008cg} and \cite{Gaiotto:2008cg,Grignani:2008is}, respectively. In fact, the $\lambda$-dependence of many other quantities like the S-matrix, the Bethe ansatz, the Zhukowsky map, the universal scaling function, etc., are also related between the AdS$_5$/CFT$_4$ and the AdS$_4$/CFT$_3$ correspondence by appropriately replacing $\lambda$ by $h(\lambda)$. Despite this fact, the subleading terms seem to be scheme dependent. 

The first indication of this scheme dependence was the observation that string theory computations \cite{McLoughlin:2008ms,Alday:2008ut,Krishnan:2008zs} of the one-loop energy shift of the spinning folded string (encoded in the universal scaling function) gave an answer that differed from the prediction of the conjectured Bethe equations \cite{Gromov:2008qe}. Two possible, but mutually exclusive, resolutions were proposed. In \cite{Gromov:2008fy}, an algebraic curve inspired regularization was used to sum the string frequencies which changed the string theory result so that it agreed with the one from the Bethe equations. Conversely, in \cite{McLoughlin:2008he}, it was shown that in the string regularization scheme, the function $h(\lambda)$ receives a one-loop correction ($a_1$ in \eqref{eqn:general-h-expansion}), and when using this contribution in the Bethe equations then the string theory result was reproduced. A similar comparison of the worldsheet and the algebraic curve computations for the circular spinning string and the analysis of different prescriptions for summing frequencies was carried out in \cite{Bandres:2009kw}. The interplay between the summation prescriptions and the constant term in the strong-coupling expansion of $h(\lambda)$ was further explored in the context of giant magnons and their dispersion relation in \cite{Abbott:2010yb}.

It does not seem that consensus has been reached in the literature as to how this puzzle should be resolved. However, it is probably wrong to attach too great importance to the function $h(\lambda)$ as it is an \emph{unphysical} object. In order to compare the results of different calculations of the same quantity, one should rather eliminate $h(\lambda)$ (resp.\ $\lambda$) from this quantity in favor of a \emph{physical reference observable} that has been computed within the same scheme.

%%%%%%%%%%%%%%%%%%%%%%%%%%%%%%%%%%%%%%%%%%%%%%%%%%%%%%%%%%%%%%%%%%%%%%%%%%%%%%%%
%%%%%%%%%%%%%%%%%%%%%%%%%%%%%%%%%%%%%%%%%%%%%%%%%%%%%%%%%%%%%%%%%%%%%%%%%%%%%%%%
\section{\texorpdfstring{$\superN=6$}{N=6} Chern-Simons matter theory}
\label{sec:ABJM-theory}

\begin{figure}%
\begin{center}
\includegraphics[scale=1]{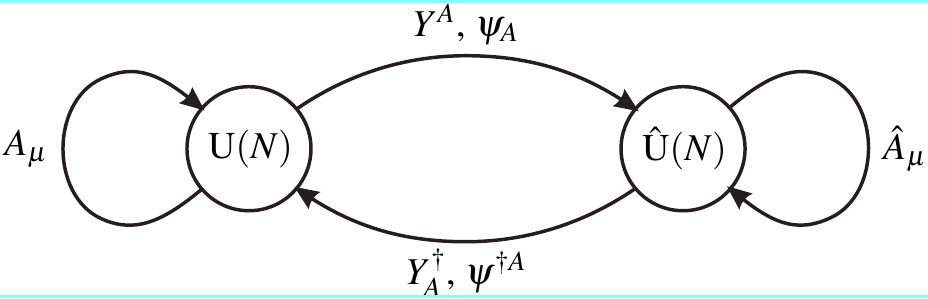}%
\end{center}
\caption{\textbf{Quiver diagram of ABJM theory.} The arrows indicate the representations of the fields under the gauge groups. The arrows are drawn from a fundamental to an anti-fundamental representation. 
% The gauge fields, $A_\mu$ and $\hat{A}_\mu$, transform in the respective adjoints, and the matter fields, $Y^A,\psi_A$ and $Y^\dagger_A,\psi^{\dagger A}$, transform in the bi-fundamentals $(\rep{N},\rep{\bar{N}})$ and $(\rep{\bar{N}},\rep{N})$, respectively. As indicated by the index $A=1..4$, the matter fields also transform under the global group $\grSU(4)_R$ in the fundamental (when the index is up) or the anti-fundamental representation (when the index is down).
}
\label{fig:ABJM-quiver}%
\end{figure}

%%%%%%%%%%%%%%%%%%%%%%%%%%%%%%%%%%%%%%%%%%%%%%%%%%%%%%%%%%%%%%%%%%%%%%%%%%%%%%%%
\paragraph{Field content.} ABJM theory is a three-dimensional superconformal Chern-Simons theory with product gauge group $\grU(N)\times\hat{\grU}(N)$ at levels $\pm k$ and specific matter content. The quiver diagram visualizing the fields of the theory and their gauge representations is drawn in \figref{fig:ABJM-quiver}. The entire field content is given by two gauge fields $A_\mu$ and $\hat{A}_\mu$, four complex scalar fields $Y^A$, and four Weyl-spinors $\psi_A$. The matter fields are $N\times N$ matrices transforming in the bi-fundamental representation of the gauge group. 

%%%%%%%%%%%%%%%%%%%%%%%%%%%%%%%%%%%%%%%%%%%%%%%%%%%%%%%%%%%%%%%%%%%%%%%%%%%%%%%%
\paragraph{Global symmetries.} The global symmetry group of ABJM theory, for Chern-Simons level%
\footnote{We are ignoring the symmetry enhancement to $\grOSp(8|4)$ at $k=1$ and $k=2$, because for the purpose of discussing integrability we have to work in the 't Hooft limit where $k$ is large.} %
%\footnote{At $k=1$ and $k=2$ the theory is expected to experience a symmetry enhancement to $\grOSp(8|4)$ \cite{Aharony:2008ug}. While a rigorous proof of this expectation is still lacking, it could be shown that the theory admits appropriate monopole operators \cite{Benna:2009xd} by means of which one could postulate the existence of additional Noether currents, which then together with the original ones generate $\grOSp(8|4)$ \cite{Gustavsson:2009pm,Kwon:2009ar}. NEW REFERENCES! \cite{Bashkirov:2010kz,Samtleben:2010eu}}%
%
$k>2$, is given by the orthosymplectic supergroup $\grOSp(6|4)$ \cite{Aharony:2008ug,Bandres:2008ry} and the ``baryonic'' $\grU(1)_b$ \cite{Aharony:2008ug}. The bosonic components of $\grOSp(6|4)$ are the R-symmetry group $\grSO(6)_R\cong\grSU(4)_R$ and the 3d conformal group $\grSp(4)\cong\grSO(2,3)$. The conformal group contains the spacetime rotations $\grSO(3)_r \cong \grSU(2)_r$ and dilatations $\grSO(2)_\Delta \cong \grU(1)_\Delta$. The fermionic part of $\grOSp(6|4)$ generates the $\superN = 6$ supersymmetry transformations. The baryonic charge $\grU(1)_b$ is $+1$ for bi-fundamental fields, $-1$ for anti-bi-fundamental fields, and $0$ for adjoint fields. The representations in which the fields transform under these symmetries are listed in \tabref{tab:ABJM-fields}. For more details about the $\grOSp(6|4)$ group theory in this context see \cite{Papathanasiou:2009en}. Finally, the model also possesses a discrete, parity-like symmetry. This might be surprising since the Chern-Simons action is not invariant but changes sign under a canonical parity transformation. The trick to make the model parity invariant is to accompany the ``naive'' parity transformation by the exchange of the two gauge group factors. The total transformation is a symmetry because the Chern-Simons terms for the two gauge group factors have opposite signs. 

\begin{table}%
\begin{center}
\begin{tabular}{c|cccccc}
              & $\grU(N)$ & $\hat{\grU}(N)$ & $\grSU(4)_{R}$ & $\grSU(2)_{r}$ & $\grU(1)_\Delta$ & $\grU(1)_b$ \\ \hline
$A_\mu$       & $\rep{N}^2$ & $\rep{1}$       & $\rep{1}$       & $\rep{3}$ & $1$ & $0$ \\
$\hat{A}_\mu$ & $\rep{1}$   & $\rep{N}^2$     & $\rep{1}$       & $\rep{3}$ & $1$ & $0$ \\
$Y^A$         & $\rep{N}$   & $\bar{\rep{N}}$ & $\rep{4}$       & $\rep{1}$ & $\half$ & $1$ \\
$\psi_{A}$    & $\rep{N}$   & $\bar{\rep{N}}$ & $\bar{\rep{4}}$ & $\rep{2}$ & $1$ & $1$ 
\end{tabular}
\end{center}
\caption{\textbf{Representations of ABJM fields.}
% The representations under the gauge group $\grU(N)\times\hat{\grU}(N)$, the R-symmetry group $\grSU(4)_{R}$, and the Lorentz group $\grSU(2)_{r}$ are given. The $\grU(1)_\Delta$ charges are the bare conformal dimensions.
}
\label{tab:ABJM-fields}
\end{table}

%%%%%%%%%%%%%%%%%%%%%%%%%%%%%%%%%%%%%%%%%%%%%%%%%%%%%%%%%%%%%%%%%%%%%%%%%%%%%%%%
\paragraph{Action.} The ABJM action was first spelled out in all detail in \cite{Benna:2008zy} in $\superN=2$ superspace and in component form. An $\superN=3$ \cite{Buchbinder:2008vi}, an $\superN=1$ \cite{Mauri:2008ai}, and an $\superN=6$ \cite{Cederwall:2008xu} superspace version is also known. The component action using the conventions of \cite{Benna:2008zy} reads
\be \label{eqn:ABJM-component-action}
  \Action \eq \frac{k}{4\pi} \int d^3x\: \Bigsbrk{ 
        \levi^{\mu\nu\lambda} \tr \bigbrk{
        A_\mu \partial_\nu A_\lambda + \tfrac{2i}{3} A_\mu A_\nu A_\lambda
        - \hat{A}_\mu \partial_\nu \hat{A}_\lambda - \tfrac{2i}{3} \hat{A}_\mu \hat{A}_\nu \hat{A}_\lambda
      }
\nl \hspace{20mm}
    - \tr (D_\mu Y)^\dagger D^\mu Y
    - i \tr \psi^\dagger \slashed{D} \psi
    - V_{\mathrm{ferm}} - V_{\mathrm{bos}}
} \; ,
\ee
where the sextic bosonic and quartic mixed potentials are
\be
   V^{\mathrm{bos}} \eq - \frac{1}{12} \tr \Bigsbrk{
          Y^A Y_A^\dagger Y^B Y_B^\dagger Y^C Y_C^\dagger 
      +   Y_A^\dagger Y^A Y_B^\dagger Y^B Y_C^\dagger Y^C
\nl\hspace{11mm}
      + 4 Y^A Y_B^\dagger Y^C Y_A^\dagger Y^B Y_C^\dagger 
      - 6 Y^A Y_B^\dagger Y^B Y_A^\dagger Y^C Y_C^\dagger 
   } \; . \\
   V^{\mathrm{ferm}} \eq \frac{i}{2} \tr \Bigsbrk{
        Y_A^\dagger Y^A \psi^{\dagger B} \psi_B
      - Y^A Y_A^\dagger \psi_B \psi^{\dagger B}
      + 2 Y^A Y_B^\dagger \psi_A \psi^{\dagger B}
      - 2 Y_A^\dagger Y^B \psi^{\dagger A} \psi_B
\nl\hspace{10mm}
      - \levi^{ABCD} Y_A^\dagger \psi_B Y_C^\dagger \psi_D
      + \levi_{ABCD} Y^A \psi^{\dagger B} Y^C \psi^{\dagger D}
      } \; .
\ee
The covariant derivative acts on bi-fundamental fields as
\be \label{eqn:cov-derivative}
  D_\mu Y = \partial_\mu Y + i A_\mu Y - i Y \hat{A}_\mu \; ,
\ee
while on anti-bi-fundamental fields it acts with $A_\mu$ and $\hat{A}_\mu$ interchanged. According to the M-theory interpretation, this theory describes the low-energy limit of $N$ M2 branes probing a $\Complex^4/\Integers_k$ singularity. The three-dimensional spacetime of ABJM theory is the world-volume of those M2 branes. For conventions and further details we refer to \cite{Benna:2008zy}.

%%%%%%%%%%%%%%%%%%%%%%%%%%%%%%%%%%%%%%%%%%%%%%%%%%%%%%%%%%%%%%%%%%%%%%%%%%%%%%%%
\paragraph{Perturbation theory and 't Hooft limit.} Note that the Chern-Simons level occurs in \eqref{eqn:ABJM-component-action} as an overall factor of the entire action. Alternatively, one can rescale the fields in such a way that all quadratic terms come without any factors of $k$ and interactions of order $n$ come with $\frac{1}{k^{n/2-1}}$. Either way, this shows that $g_{\mathrm{CS}}^2 \equiv \frac{1}{k}$ acts like a coupling constant of ABJM theory, quite similar to $g_{\mathrm{YM}}^2$ in $\superN=4$ SYM, though of course $k$ has to be an integer to preserve non-abelian gauge symmetry. As announced in the introduction, the theory can be restricted to the planar sector by taking the 't Hooft limit \eqref{eqn:tHooft} which introduces the effective coupling
\be \label{eqn:ABJM-coupling}
  \lambda \equiv g_{\mathrm{CS}}^2 N = \frac{N}{k} \; .
\ee
In this limit the theory becomes integrable \cite{Minahan:2008hf} (see also \cite{Gaiotto:2008cg,Bak:2008cp}) in the same sense as we are used to in planar $\superN=4$ SYM theory and as we will discuss below.

%%%%%%%%%%%%%%%%%%%%%%%%%%%%%%%%%%%%%%%%%%%%%%%%%%%%%%%%%%%%%%%%%%%%%%%%%%%%%%%%
\paragraph{Gauge group.} The model can be generalized to have gauge group $\grU(M)_{k}\times\grU(N)_{-k}$ \cite{Aharony:2008gk}. This generalization goes by the name ABJ theory. The M-theory interpretation is given by $\min(M,N)$ M2 branes allowed to move freely on $\Complex^4/\Integers_k$ and $\abs{M-N}$ fractional M2 branes stuck to the singularity. The gauge theory action is formally the same as in \eqref{eqn:ABJM-component-action}, except that the matter fields are now given by rectangular matrices. Thus two 't Hooft couplings can be defined by
\be \label{eqn:ABJ-coupling}
  \lambda = \frac{M}{k} \comma \hat{\lambda} = \frac{N}{k} \; ,
\ee
and it becomes possible to take different planar limits depending on the ratio of $\lambda$ and $\hat{\lambda}$. On the other hand, the generalized parity invariance of the ABJM theory is explicitly broken, because now the two gauge group factors cannot be exchanged anymore.

%%%%%%%%%%%%%%%%%%%%%%%%%%%%%%%%%%%%%%%%%%%%%%%%%%%%%%%%%%%%%%%%%%%%%%%%%%%%%%%%
\paragraph{Deformation.} It is possible to introduce independent Chern-Simons levels $k$ and $\hat{k}$ for the two gauge groups $\grU(N)$ and $\hat{\grU}(N)$ that do not sum to zero. This generalized theory possesses less supersymmetry and less global symmetry. It is proposed to be dual to a type IIA background with the Romans mass $F_0 = k + \hat{k}$ turned on \cite{Gaiotto:2009mv}. This modification, however, seems to break integrability \cite{Forcella:2009gm}.

%%%%%%%%%%%%%%%%%%%%%%%%%%%%%%%%%%%%%%%%%%%%%%%%%%%%%%%%%%%%%%%%%%%%%%%%%%%%%%%%
%%%%%%%%%%%%%%%%%%%%%%%%%%%%%%%%%%%%%%%%%%%%%%%%%%%%%%%%%%%%%%%%%%%%%%%%%%%%%%%%
\section{From ABJM theory to the integrable model}
\label{sec:ABJMtoInt}

%%%%%%%%%%%%%%%%%%%%%%%%%%%%%%%%%%%%%%%%%%%%%%%%%%%%%%%%%%%%%%%%%%%%%%%%%%%%%%%%
\paragraph{Spin-chain picture.} The integrability of the planar ABJM theory is best described in terms of an integrable $\grOSp(6|4)$ spin-chain which represents single trace operators \cite{Minahan:2008hf}. A qualitative difference between the case at hand and the case of $\superN=4$ SYM is that the ABJM spin-chain is an ``alternating spin-chain.'' Because the matter fields are in bi-fundamental representations of the product gauge group $\grU(N)\times\hat{\grU}(N)$, gauge invariant operators need to be built from products of fields that transform alternatingly in the representations $(\rep{N},\rep{\bar{N}})$ and $(\rep{\bar{N}},\rep{N})$, e.g.
\be \label{eqn:spin-chain-vacuum}
  \tr( Y^1 Y_4^\dagger Y^1 Y_4^\dagger \cdots ) \; .
\ee
Thus, the spin-chain has even length and the fields on the odd sites are distinct from the ones on the even sites. On the odd sites, we can have any of the 4$_B$+8$_F$ fields $Y^A$, $\psi_{A\alpha}$, and on the even sites, we can have any of the 4$_B$+8$_F$ fields $Y^\dagger_A$, $\psi^{\dagger A}_\alpha$. We can also act with an arbitrary number of derivatives $D_\mu = D_{\alpha\beta}$ onto the fields, but derivatives do not introduce extra sites. Also field strength insertions do not count as extra sites as they can be written as anti-symmetrized derivatives.

%%%%%%%%%%%%%%%%%%%%%%%%%%%%%%%%%%%%%%%%%%%%%%%%%%%%%%%%%%%%%%%%%%%%%%%%%%%%%%%%
\paragraph{Spin-chain excitations.} In the spin-chain description, the ABJM fields are distinguished according to whether they represent the vacuum (or ``down spin''), or elementary or multiple excitations. A convenient and common choice for the vacuum spin-chain is the BPS operator \eqref{eqn:spin-chain-vacuum}, i.e. $Y^1$ is the vacuum on the odd sites, and $Y_4^\dagger$ is the vacuum on the even sites.

Selecting a vacuum breaks the $\grOSp(6|4)$ symmetry group of ABJM theory down to $\grSU(2|2)\times\grU(1)_{\mathrm{extra}}$ which becomes the symmetry group of the spin-chain model \cite{Minahan:2008hf,Gaiotto:2008cg}. The bosonic components of this $\grSU(2|2)$ are $\grSU(2)_G \times \grSU(2)_r \times \grU(1)_E$, where $\grSU(2)_G$ is the unbroken part of $\grSU(4)_R$, $\grSU(2)_r \cong \grSO(1,2)_r$ is the Lorentz group, and $\grU(1)_E$ is the spin-chain energy $E=\Delta - J$ which itself is a combination of the conformal dimension $\Delta$ and a broken $\grSU(4)_R$ generator $J$. The charges of the fields under these groups are listed and explained in \tabref{tab:ABJM-spin-chain-charges}.

\begin{table}%
\begin{center}
%\begin{footnotesize}
\begin{scriptsize}
\begin{tabular}{l|c|ccc|cc|c}
                     & $\grSU(4)_R$ & $\grSU(2)_{G'}$ & $\grSU(2)_{G}$ & $\grU(1)_{\mathrm{extra}}$ & $\grU(1)_\Delta$ & $\grSU(2)_r$ & $\grU(1)_E$ \\
                     & $[p_1,q,p_2]$ & $J$ & $$ & $$ & $\Delta$ & $s$ & $E=\Delta-J$  \\ \hline
$Y^1$                & $[\,1\,,\,0\,,\,0\,]$ & $+1/2$ & 0      & $+1$ & $1/2$ & $0$      & $0$   \\
$Y^2$                & $[-1,1,0]$            & 0      & $+1/2$ & $-1$ & $1/2$ & $0$      & $1/2$ \\
$Y^3$                & $[0,-1,1]$            & 0      & $-1/2$ & $-1$ & $1/2$ & $0$      & $1/2$ \\
$Y^4$                & $[0,0,-1]$            & $-1/2$ & 0      & $+1$ & $1/2$ & $0$      & $1$   \\ \hline
$\psi_{1\pm}$        & $[-1,0,0]$            & $-1/2$ & 0      & $-1$ & $1$   & $\pm1/2$ & $3/2$ \\
$\psi_{2\pm}$        & $[1,-1,0]$            & 0      & $-1/2$ & $+1$ & $1$   & $\pm1/2$ & $1$   \\
$\psi_{3\pm}$        & $[0,1,-1]$            & 0      & $+1/2$ & $+1$ & $1$   & $\pm1/2$ & $1$   \\
$\psi_{4\pm}$        & $[\,0\,,\,0\,,\,1\,]$ & $+1/2$ & 0      & $-1$ & $1$   & $\pm1/2$ & $1/2$ \\ \hline
$D_{0}$              & $[\,0\,,\,0\,,\,0\,]$ & $0$    & $0$    & $0$  & $1$   & $0$      & $1$   \\
$D_{\pm}$            & $[\,0\,,\,0\,,\,0\,]$ & $0$    & $0$    & $0$  & $1$   & $\pm1$   & $1$   \\ \hline
$Y^\dagger_1$        & $[-1,0,0]$            & $-1/2$ & 0      & $-1$ & $1/2$ & $0$      & $1$   \\
$Y^\dagger_2$        & $[1,-1,0]$            & 0      & $-1/2$ & $+1$ & $1/2$ & $0$      & $1/2$ \\
$Y^\dagger_3$        & $[0,1,-1]$            & 0      & $+1/2$ & $+1$ & $1/2$ & $0$      & $1/2$ \\
$Y^\dagger_4$        & $[\,0\,,\,0\,,\,1\,]$ & $+1/2$ & 0      & $-1$ & $1/2$ & $0$      & $0$   \\ \hline
$\psi^{\dagger1\pm}$ & $[\,1\,,\,0\,,\,0\,]$ & $+1/2$ & 0      & $+1$ & $1$   & $\pm1/2$ & $1/2$ \\
$\psi^{\dagger2\pm}$ & $[-1,1,0]$            & 0      & $+1/2$ & $-1$ & $1$   & $\pm1/2$ & $1$   \\
$\psi^{\dagger3\pm}$ & $[0,-1,1]$            & 0      & $-1/2$ & $-1$ & $1$   & $\pm1/2$ & $1$   \\
$\psi^{\dagger4\pm}$ & $[0,0,-1]$            & $-1/2$ & 0      & $+1$ & $1$   & $\pm1/2$ & $3/2$
\end{tabular}
\end{scriptsize}
%\end{footnotesize}
\end{center}
\caption{\textbf{Charges of fields.} The R-symmetry group $\grSO(6)_R\cong\grSU(4)_R$ splits up into $\grSU(2)_{G'} \times \grSU(2)_G \times \grU(1)_{\mathrm{extra}}$, and the conformal group $\grSp(2,2)\cong\grSO(2,3)$ splits up into $\grU(1)_\Delta \times \grSU(2)_r$. The symmetry group of the spin-chain is $\grSU(2|2) \times \grU(1)_{\mathrm{extra}} \supset \grSU(2)_G \times \grSU(2)_r \times \grU(1)_E \times \grU(1)_{\mathrm{extra}}$. The $\grU(1)_J$ generator $J = \frac{p_1+q+p_2}{2}$ is the Cartan generator of $\grSU(2)_{G'}$, and the $\grU(1)_E$ generator $E$ is given by the difference $\Delta-J$.}
\label{tab:ABJM-spin-chain-charges}
\end{table}

By construction, the ground state spin-chain \eqref{eqn:spin-chain-vacuum} has energy $E = \Delta - J = 0$. This spin-chain can be excited by replacing one of the vacuum fields by a different field or by acting with a covariant derivative. This procedure increases the energy in quanta of $\delta E = 1/2$ by a total amount that can be read off from the last column in \tabref{tab:ABJM-spin-chain-charges}. If the energy increases by $1/2$, then the excitation is an elementary one. We find that the elementary excitations on the odd and even sites are given by
\begin{subequations} \label{eqn:spin-chain-particles}
\be
\mbox{``$A$''-particles:} && (Y^2,Y^3|\psi_{4+},\psi_{4-}) \; , \label{eqn:spin-chain-A-particle} \\[1mm]
\mbox{``$B$''-particles:} && (Y^\dagger_3,Y^\dagger_2|\psi^{\dagger1}_+,\psi^{\dagger1}_-) \; , \label{eqn:spin-chain-B-particle}
\ee
\end{subequations}
respectively \cite{Gaiotto:2008cg}. These are the two multiplets anticipated in \eqref{eqn:A+B}. All other fields correspond to composite excitations and are listed in \tabref{tab:multi-excitations}.

\begin{table}%
\begin{center}
\begin{tabular}{l|l|l}
& Multi-excitation & made of \\ \hline
Double excitations & $Y^\dagger_1 {\gray Y^1}    \:,\; Y^4 {\gray Y^\dagger_4}$      & $Y^2 Y^\dagger_2 \pm Y^3 Y^\dagger_3$ \\
                   & $\psi_2 {\gray Y^\dagger_4} \:,\; \psi^{\dagger3} {\gray Y^1}$  & $\psi_4 Y^\dagger_2 \pm Y^3 \psi^{\dagger1}$ \\
                   & $\psi_3 {\gray Y^\dagger_4} \:,\; \psi^{\dagger2} {\gray Y^1}$  & $\psi_4 Y^\dagger_3 \pm Y^2 \psi^{\dagger1}$ \\ \hline
Triple excitations & $\psi_1 {\gray Y^\dagger_4 Y^1}$                                & $Y^2 \psi^{\dagger1} Y^3$           \\
                   & $\psi^{\dagger4} {\gray Y^1 Y^\dagger_4}$                       & $Y^\dagger_2 \psi_4 Y^\dagger_3$    \\
                   & $D_\mu {\gray Y^1 Y^\dagger_4}$                                 & $\psi_4 \gamma_\mu \psi^{\dagger1}$
\end{tabular}
\end{center}
\caption{\textbf{Multi-excitations.} In order to determine which elementary excitations a composite is made out of, one needs to compare their $\grSU(2|2) \times \grU(1)_{\mathrm{extra}}$ charges. E.g. for the triple excitation $\psi_1$ one can check that the charges of $\psi_1$ together with the two background fields $Y^1 Y^\dagger_4$ coincide with the charges of the three elementary excitations $Y^2 \psi^{\dagger1} Y^3$.} 
\label{tab:multi-excitations}
\end{table}

%%%%%%%%%%%%%%%%%%%%%%%%%%%%%%%%%%%%%%%%%%%%%%%%%%%%%%%%%%%%%%%%%%%%%%%%%%%%%%%%
\paragraph{Subsectors.} A subsector is a set of fields which is closed under the action of the spin-chain Hamiltonian, i.e. there is no overlap between spin-chains from within a subsector with spin-chains from outside. The subsectors of ABJM theory above the vacuum \eqref{eqn:spin-chain-vacuum} are listed in \tabref{tab:subsectors}. To prove that these sectors are closed to all orders in perturbation theory, one defines a positive semi-definite charge $P = n_1 p_1 + n_2 q + n_3 p_2 + n_4 \Delta + n_5 s + n_6 b \ge 0$ from the eigenvalues of all operators that commute with the spin-chain Hamiltonian $E=\Delta - J$. These are the 5 Cartan generators of $\grOSp(6|4)$ and the baryonic charge $U(1)_b$. The set of fields with $P=0$ constitute a closed subsector. Different subsectors are obtained by different choices for the numbers $n_i$.

\begin{table}%
\begin{center}
\begin{tabular}{l|lll}
Subsector & Vacuum & Single & Double \\ \hline
%$P = -p_1-q-p_2+2\Delta$:
Vacuum & $Y^1$ $Y^\dagger_4$ \\
%$P = -p_1-2q-p_2+2\Delta$:
$\grSU(2)\times\grSU(2)$ & $Y^1$ $Y^\dagger_4$ & $Y^2$ $Y^\dagger_3$ \\
%$P = -p_1-q-p_2+2\Delta-2s$:
$\grOSp(2|2)$ & $Y^1$ $Y^\dagger_4$ & $\psi_{4+}$ $\psi^{\dagger1}_+$ & $D_+$\\
%$P = -p_1-2q-p_2+2\Delta-2s$:
$\grOSp(4|2)$ & $Y^1$ $Y^\dagger_4$ & $Y^2$ $\psi_{4+}$ $Y^\dagger_3$ $\psi^{\dagger1}_+$ & $D_+$ $\psi_{3+}$ $\psi^{\dagger2}_+$ \\ \hline
%$P = 0$:
%$\grOSp(6|4)$ & all fields \\
$\grSU(2)$ & $Y^1$ $Y^\dagger_4$ & $Y^2$ \\ 
$\grSU(1|1)$ & $Y^1$ $Y^\dagger_4$ & $\psi_{4+}$ \\ 
$\grSU(2|1)$ & $Y^1$ $Y^\dagger_4$ & $Y^2$ $\psi_{4+}$ \\ 
$\grSU(3|2)$ & $Y^1$ $Y^\dagger_4$ & $Y^2$ $Y^3$ $\psi_{4+}$ $\psi_{4-}$
\end{tabular}
\end{center}
\caption{\textbf{Subsectors.} This list of closed subsectors above the vacuum $\tr( Y^1 Y_4^\dagger Y^1 Y_4^\dagger \cdots )$ is complete, although a specific subsector can be realized also by other fields. That would correspond to a different embedding of the sector into the full theory. Note that there is no closed $\grSL(2)$ sector that is made only out of derivatives as we had in $\superN=4$ SYM. This is because derivatives are double excitations of fermions with the above choice of vacuum. However, it is also possible to consider closed subsectors based on a different vacuum. There is, for instance, an $\grSL(2)$ sector built from derivatives onto the vacuum $\tr(Y^1\psi^{\dagger 1})^L$ \cite{Zwiebel:2009vb}, which was studied e.g.\ in \cite{Beccaria:2009ny,Beccaria:2009wb}.}
\label{tab:subsectors}
\end{table}

%%%%%%%%%%%%%%%%%%%%%%%%%%%%%%%%%%%%%%%%%%%%%%%%%%%%%%%%%%%%%%%%%%%%%%%%%%%%%%%%
\paragraph{Spin-chain Hamiltonian.} Various works have computed the spin-chain Hamiltonian for different subsectors to different loop orders with different methods in different approximations. The first results were obtained in the $\grSU(4)$ sector\footnote{This sector is closed at two-loop order but not beyond.} at two\footnote{There is no contribution to the Hamiltonian at an odd number of loops as in three dimensions no such Feynman diagram is logarithmically divergent.} loops \cite{Minahan:2008hf,Bak:2008cp} where the spin-chain Hamiltonian reads
\be
  H = \frac{\lambda^2}{2} \sum_{l=1}^{2L} \bigbrk{ 2 - 2 P_{l,l+2} + P_{l,l+2} K_{l,l+1} + K_{l,l+1} P_{l,l+2} } \; .
\ee
with $P_{l,m}$ and $K_{l,m}$ being the permutation and the trace operator, respectively, and $2L$ being the length of the spin-chain. This Hamiltonian has been proven to be integrable by means of an algebraic Bethe ansatz \cite{Minahan:2008hf,Bak:2008cp}. In the $\grSU(2)\times\grSU(2)$ sector, independently studied in \cite{Gaiotto:2008cg}, the trace operators annihilate the states and the Hamiltonian reduces to the sum of two decoupled Heisenberg XXX$_{1/2}$ Hamiltonians, one acting onto the even sites and one acting onto the odd sites. The only coupling between these two sublattices comes from the cyclicity condition which says that the \emph{total} momentum of all excitations has to be zero (mod $2\pi$), not individually for the even and odd sites. Nevertheless, the Hamiltonians will continue to be decoupled up to six loop order \cite{Gromov:2008qe}.

The extension of the two-loop Hamiltonian to the full theory was derived in \cite{Zwiebel:2009vb} and \cite{Minahan:2009te}. The integrability in the $\grOSp(4|2)$ sector was proved by means of a Yangian construction \cite{Zwiebel:2009vb}. 
%by imposing the algebraic constraints from supersymmetry and the restrictions that arise from the planarity of the underlying Feynman diagrams on the possible form of the Hamiltonian, and in \cite{Minahan:2009te} from ...
The generalization to ABJ theory at two loops was studied in the scalar sector \cite{Bak:2008vd} and the full theory \cite{Minahan:2009te}, which at this perturbative order amounts to replacing $\lambda^2$ in the ABJM result by $\lambda\hat{\lambda}$, cf.\ \eqref{eqn:ABJ-coupling}. That means that the absence of parity in ABJ theory is not visible at two loop order.

Beyond two loops only the dispersion relation, i.e.\ the eigenvalue of the Hamiltonian on spin-chains with a single excitation, is known to date. It is of the general form \eqref{eqn:general-disp-rel}. The expansion of the interpolating function $h$ to four-loop order was computed for the ABJM and the ABJ theory in \cite{Minahan:2009aq,Minahan:2009wg,Leoni:2010tb} with the result
\be \label{eqn:h-func-ABJ}
  h^2(\lambda,\hat{\lambda}) = \lambda\hat{\lambda} - (\lambda\hat{\lambda})^2 \Biggsbrk{
    \frac{2\pi^2}{3} + \frac{\pi^2}{6} \biggbrk{\frac{\lambda-\hat{\lambda}}{\sqrt{\lambda\hat{\lambda}}}}^2 
  } \; ,
\ee
where the ABJM expression is obtained from this by setting the two 't Hooft couplings equal to each other. We see that $h(\lambda,\lambda)$ is for the form \eqref{eqn:general-h-expansion} with $c_1=-\pi^2/3$. Note that \eqref{eqn:h-func-ABJ} is invariant under the exchange of $\lambda$ and $\hat{\lambda}$, even though ABJ theory lacks manifest parity invariance. The fact that parity is not broken in the spin-chain picture is \emph{not} a consequence of integrability, because as shown in \cite{Bak:2008vd} there are integrable but parity breaking spin-chain Hamiltonians already at two loops. Alternative explanations for the non-visibility of parity breaking were proposed \cite{Bak:2008vd}. In ABJ theory one can also study the limit $\lambda \gg \hat{\lambda}$ \cite{Minahan:2009aq}. In this limit, the Hamiltonian of the $\grSU(2)\times\grSU(2)$ sector is, at any loop order, proportional to two decoupled Heisenberg spin-chain Hamiltonians \cite{Minahan:2009aq}. An exact expression for the $\lambda$-dependent prefactor, which gives a prediction for the function $h(\lambda,\hat{\lambda})$ in the limit $\hat{\lambda} \ll \lambda$, has been conjectured in \cite{Minahan:2010nn}. Very recently, even for the case when $\lambda=\hat{\lambda}$ an all-order guess for $h^2(\lambda)$ was made \cite{Leoni:2010tb}, that is in line with the weak and strong coupling data.

At six loops only a subset of Feynman diagrams have been evaluated, namely those which move the impurities along the spin-chain by the maximal amount that is possible at this loop order \cite{Bak:2009tq}. The contributions from this subset to the dilatation operator are consistent with the corresponding spin-chain being integrable \cite{Bak:2009tq}.

Also non-planar contributions to the two-loop dilatation operator have been computed in the $\grSU(2)\times\grSU(2)$ sector \cite{Kristjansen:2008ib}. The degeneracy of the dimensions of parity pairs at the planar level, which is a signature of integrability, is lifted by the non-planar contributions \cite{Kristjansen:2008ib}. At the non-planar level, one can also observe the breaking of parity in the ABJ theory already at two loops \cite{Caputa:2009ug}.

%%%%%%%%%%%%%%%%%%%%%%%%%%%%%%%%%%%%%%%%%%%%%%%%%%%%%%%%%%%%%%%%%%%%%%%%%%%%%%%%
%%%%%%%%%%%%%%%%%%%%%%%%%%%%%%%%%%%%%%%%%%%%%%%%%%%%%%%%%%%%%%%%%%%%%%%%%%%%%%%%
\section{Superstrings on \texorpdfstring{$\AdS_4 \times \CP^3$}{AdS4xCP3}}
\label{sec:IIA-theory}

% IIA opposite chirality
% IIB equal chirality

%%%%%%%%%%%%%%%%%%%%%%%%%%%%%%%%%%%%%%%%%%%%%%%%%%%%%%%%%%%%%%%%%%%%%%%%%%%%%%%%
\paragraph{String background.} $\AdS_4 \times \CP^3$ with two- and four-form fluxes turned on is a solution to IIA supergravity that preserves 24 out of 32 supersymmetries \cite{Nilsson:1984bj}, i.e.\ unlike $\AdS_5 \times \Sphere^5$ it is not maximally supersymmetric. The $\AdS_4 \times \CP^3$ superspace geometry has been constructed in \cite{Gomis:2008jt}. The fermionic coordinates $\Theta^{1..32} = \bigbrk{\vartheta^{1..24},\upsilon^{1..8} }$ split into 24 coordinates $\vartheta$, which correspond to the unbroken supersymmetries of the background, and eight coordinates $\upsilon$ corresponding to the broken supersymmetries.

%%%%%%%%%%%%%%%%%%%%%%%%%%%%%%%%%%%%%%%%%%%%%%%%%%%%%%%%%%%%%%%%%%%%%%%%%%%%%%%%
\paragraph{Green-Schwarz action.} Although formal expressions for the Green-Schwarz superstring action exist for any type II supergravity background \cite{Grisaru:1985fv}, in practice it is generically hopeless to find exact expressions for the supervielbeins. Nevertheless, utilizing the connection to M-theory on $\AdS_4 \times \Sphere^7$, all functions that are required to write down the Nambu-Goto form of the action, in particular the supervielbeins and the NS-NS two-form superfield, were explicitly spelled out in \cite{Gomis:2008jt}. Two different $\kappa$-gauge-fixed versions of the action were given in \cite{Grassi:2009yj} and \cite{Uvarov:2009hf}. The latter version was obtained by a double dimensional reduction of the action of the supermembrane on $\AdS_4\times\Sphere^7$.

%%%%%%%%%%%%%%%%%%%%%%%%%%%%%%%%%%%%%%%%%%%%%%%%%%%%%%%%%%%%%%%%%%%%%%%%%%%%%%%%
\paragraph{Coset action.} A less complete, but sometimes more pragmatic, approach to strings on $\AdS_4 \times \CP^3$ has earlier been taken in \cite{Arutyunov:2008if} and \cite{Stefanski:2008ik}. The observation is that $\AdS_4$ is the coset $\grSO(2,3)/\grSO(1,3)$ and $\CP^3$ is the coset $\grSO(6)/\grU(3)$, and that $\grSO(2,3)\times\grSO(6)$ is the bosonic subgroup of $\grOSp(6|4)$. Thus the idea is to write the superstrings action as a sigma-model on the supercoset
\be \label{eqn:osp-coset}
  \frac{\grOSp(6|4)}{\grSO(1,3)\times \grU(3)} \; ,
\ee
analogously to the $\grPSU(2,2|4)/\grSO(1,4)\times\grSO(5)$ coset model for superstrings on $\AdS_5 \times \Sphere^5$ \cite{Metsaev:1998it}, which itself was inspired by the WZW-type action for strings in flat space \cite{Henneaux:1984mh}. Again it is possible to define a $\Integers_4$ grading \cite{Berkovits:1999zq} of the (complexified) algebra \cite{Arutyunov:2008if,Stefanski:2008ik}, and when this grading is used to split up the current one-form $A = -g^{-1}dg = A^{(0)} + A^{(1)} + A^{(2)} + A^{(3)}$, constructed from a parametrization of the coset representatives $g$, then the coset action is given by
\be \label{eqn:cosetaction}
  \Action = - \frac{R^2}{4\pi\alpha'} \int\!d\sigma\,d\tau \:
            \str \Bigsbrk{ \sqrt{-h}\, h^{\alpha\beta} \, A^{(2)}_\alpha A^{(2)}_\beta 
            + \kappa \levi^{\alpha\beta} \, A^{(1)}_\alpha A^{(3)}_\beta } \; .
\ee
The explicit form of this sigma-model action can look quite differently depending on the choice of coset representative and the choice of gauge \cite{Arutyunov:2008if,Stefanski:2008ik,Uvarov:2008yi,Zarembo:2009au}.

%%%%%%%%%%%%%%%%%%%%%%%%%%%%%%%%%%%%%%%%%%%%%%%%%%%%%%%%%%%%%%%%%%%%%%%%%%%%%%%%
\paragraph{Fermions, $\kappa$-symmetry and singular configurations.} There is a subtle problem with the coset action \eqref{eqn:cosetaction}. The supercoset \eqref{eqn:osp-coset} has only 24 fermionic directions, which is the number of supersymmetries preserved by the background. However, independent of how many supersymmetries are preserved, the Green-Schwarz superstring always requires two Majorana-Weyl fermions with a total number of 32 degrees for freedom. Thus the coset model misses 8 fermions and can therefore not be equivalent to the GS string! This problem did not exist in the case of $\AdS_5 \times \Sphere^5$ because that background is maximally supersymmetric and the corresponding supercoset has 32 fermionic directions.

It has been argued that the eight missing fermions $\upsilon$ are part of the 16 fermionic degrees of freedom that due to $\kappa$-gauge symmetry are unphysical anyway, i.e.\ to think of the coset action on \eqref{eqn:osp-coset} as an action with $\kappa$-symmetry partially gauge-fixed. Of the remaining 24 fermions $\vartheta$, further 8 should then be unphysical. For this interpretation to be correct, the rank of $\kappa$-symmetry of the coset action must be 8. This is in fact true for generic bosonic configurations \cite{Arutyunov:2008if,Stefanski:2008ik}, unfortunately however not for strings that move only in the AdS part of the background, in which case the rank of $\kappa$-symmetry is 12 \cite{Arutyunov:2008if}. This means that on such a ``singular configuration'' the coset model is a truncation of the GS string where instead of removing 8 unphysical fermions (from 32 to 24), 4 physical fermions have been put to zero, while 4 unphysical fermions have been retained. A similarly singular configuration from the point of view of the coset model is given by the worldsheet instanton in $\CP^3$ of the Wick-rotated theory \cite{Cagnazzo:2009zh}.

The upshot is that the coset model is generically equivalent to the GS string, but \emph{not} on singular backgrounds. The consequence is that these singular backgrounds cannot be quantized semi-classically within the coset description.

%%%%%%%%%%%%%%%%%%%%%%%%%%%%%%%%%%%%%%%%%%%%%%%%%%%%%%%%%%%%%%%%%%%%%%%%%%%%%%%%
\paragraph{Near plane-wave expansion.} One method for dealing with a curved RR-background at the quantum level is to take a Penrose limit of the geometry which leads to a solvable plane-wave background and then to include curvature corrections perturbatively. Penrose limits of the $\AdS_4\times\CP^3$ background were studied in \cite{Nishioka:2008gz,Gaiotto:2008cg,Grignani:2008is,Astolfi:2009qh,Grignani:2009ny}. The near plane-wave Hamiltonian was derived in a truncation\footnote{This truncation is not consistent and the absence of the fermions yields divergences, which were regularized using $\zeta$-function regularization. Up to so-called ``non-analytic'' terms, the result is correct.} to the bosonic sector in \cite{Astolfi:2008ji,Sundin:2008vt}, for a sector including fermions in \cite{Sundin:2009zu}, and for the full theory in \cite{Astolfi:2009qh}. Taking the Penrose limit of the $\AdS_4 \times \CP^3$ geometry in the AdS-light-cone gauge \cite{Uvarov:2009hf}, one ends up with a trivial plane-wave, namely flat space \cite{Uvarov:2009nk}. In this case, not just the near-flat-space but the exact Hamiltonian is known \cite{Uvarov:2009nk}.

%%%%%%%%%%%%%%%%%%%%%%%%%%%%%%%%%%%%%%%%%%%%%%%%%%%%%%%%%%%%%%%%%%%%%%%%%%%%%%%%
\paragraph{Pure spinors.} The pure spinor formulation of the superstring on $\AdS_4\times\CP^3$ was developed in \cite{Fre:2008qc,Bonelli:2008us,D'Auria:2008cw}. This approach is suitable for the covariant quantization of the string.

%%%%%%%%%%%%%%%%%%%%%%%%%%%%%%%%%%%%%%%%%%%%%%%%%%%%%%%%%%%%%%%%%%%%%%%%%%%%%%%%
%%%%%%%%%%%%%%%%%%%%%%%%%%%%%%%%%%%%%%%%%%%%%%%%%%%%%%%%%%%%%%%%%%%%%%%%%%%%%%%%
\section{From \texorpdfstring{$\AdS_4 \times \CP^3$}{AdS4xCP3} to the integrable model}
\label{sec:IIAtoInt}

%%%%%%%%%%%%%%%%%%%%%%%%%%%%%%%%%%%%%%%%%%%%%%%%%%%%%%%%%%%%%%%%%%%%%%%%%%%%%%%%
\paragraph{Evidence for integrability.} The purely bosonic sigma-model on $\AdS_4 \times \CP^3$ is integrable at the classical level, though quantum corrections spoil the integrability \cite{Abdalla:1982yd,Abdalla:1984en}. For the super-coset model, classical integrability is also proven \cite{Arutyunov:2008if,Stefanski:2008ik}. The Lax connection found in \cite{Bena:2003wd} for the $\AdS_5\times\Sphere^5$ case as a means of writing the equations of motion in a manifestly integrable form is directly applicable here. Moreover, the absence of particle production in the coset sigma-model has been shown explicitly for bosonic amplitudes at tree-level \cite{Kalousios:2009ey}. However, we know that the full GS string is more than the coset model. Evidence for the classical integrability of the complete $\AdS_4\times\CP^3$ superstring was recently given in \cite{Sorokin:2010wn} by constructing a Lax connection that was shown to be flat for (a) strings that move in a certain subspace that is different from the coset model and (b) the full theory to at least second order in fermions. Different integrable reductions of the sigma model have also been studied \cite{Ahn:2008hj,Rashkov:2008rm,Dukalski:2009pr}.

%%%%%%%%%%%%%%%%%%%%%%%%%%%%%%%%%%%%%%%%%%%%%%%%%%%%%%%%%%%%%%%%%%%%%%%%%%%%%%%%
\paragraph{Matching $\mathbf{AdS_4\times CP^3}$ to ABJM theory.} The metric on $\AdS_4\times \CP^3$ has the two factors
\be \label{eqn:metric-AdS4CP3}
  ds^2 = R^2 \Bigsbrk{ \tfrac{1}{4} ds^2_{\AdS_4} + ds^2_{\CP^3}} \; ,
\ee
where $R$ is the radius of $\CP^3$ which is twice the radius of $\AdS_4$. This relative size is demanded by supersymmetry and comes out automatically when one starts from the coset action \eqref{eqn:cosetaction}. The radius $R$ is related to the 't Hooft coupling $\lambda$ of ABJM theory by \eqref{eqn:stringparameter}. In global coordinates the metric for $\AdS_4$ reads
\be \label{eqn:metric-AdS4}
  ds^2_{\AdS_4} = -\cosh^2\rho \, dt^2 + d\rho^2 + \sinh^2\rho \bigbrk{ d\theta^2 + \sin^2\theta\, d\varphi^2}
\ee
with coordinate ranges $\rho=0\ldots\infty$, $t=-\infty\ldots\infty$, $\theta=0\ldots\pi$, and $\varphi=0\ldots2\pi$. The metric on $\CP^3$ is the standard Fubini-Study metric and can be written as
\be \label{eqn:metric-CP3}
  ds^2_{\CP^3} \eq d\xi^2 + \cos^2\xi\sin^2\xi \Bigsbrk{ d\psi + \half\cos\theta_1\,d\varphi_1 - \half \cos\theta_2\,d\varphi_2 }^2 \nl
                   + \quarter \cos^2\xi \Bigsbrk{ d\theta_1^2 + \sin^2\theta_1 \, d\varphi_1^2 }
                   + \quarter \sin^2\xi \Bigsbrk{ d\theta_2^2 + \sin^2\theta_2 \, d\varphi_2^2 } \; .
\ee
The coordinates $(\theta_1,\varphi_1)$ and $(\theta_2,\varphi_2)$ parameterize two two-spheres, the angle $\xi=0\ldots\frac{\pi}{2}$ determines their radii, and the angle $\psi=0\ldots2\pi$ corresponds to the $\grU(1)_R$ isometry.

% This table belongs further down but I have it to appear up here.
\begin{table}
\begin{center}
\begin{tabular}{l|l|l}
  field & mass & dispersion relation \\ \hline
  $t$, $\psi$                     & $0$        & $\omega_n = n$ \\
  $x_{1,2,3}$, $\xi$              & $\kappa$   & $\omega_n = \sqrt{\kappa^2 + n^2}$ \\
  $\theta_{1,2}$, $\varphi_{1,2}$ & $\kappa/2$ & $\omega_n = \sqrt{(\kappa/2)^2 + n^2} \pm \kappa/2$
\end{tabular}
\end{center}
\caption{\textbf{Spectrum of fluctuations about the point-like string.} Two linear combinations of $\theta_{1,2}$ and $\varphi_{1,2}$ possess the dispersion relation with $+\kappa/2$, and two other linear combinations the one with $-\kappa/2$.}
\label{tab:fluctuations-point}
\end{table}

The background admits five Killing vectors
\be \label{eqn:killing-vectors}
  E = -i \partial_t
  \comma
  S = -i \partial_\varphi
  \comma
  J_{\varphi_1} = -i \partial_{\varphi_1}
  \comma
  J_{\varphi_2} = -i \partial_{\varphi_2}
  \comma
  J_{\psi} = -i \partial_\psi
\ee
leading to the five conserved charges: the worldsheet energy $E$, the AdS-spin $S$ and the $\CP^3$ momenta $J_{\varphi_1}$, $J_{\varphi_2}$, and $J_\psi$. Note that this is one conserved charge less than in the $\AdS_5 \times \Sphere^5$ case where there are two AdS-spins. This shows that $\AdS_4\times\CP^3$ is less symmetric. The charges \eqref{eqn:killing-vectors} are one choice of Cartan generators of $\grSO(3,2)\times\grSU(4)$. The angular momenta $J_{\varphi_1}$ and $J_{\varphi_2}$ correspond to the Cartan generators of two $\grSU(2)$ subgroups that on the gauge theory side transform $(Y^1,Y^2)$ and $(Y^3,Y^4)$, respectively. The angular momentum $J_\psi$ is the $\grU(1)_R$ generator. Thus, the angular momenta are related to the charges in \tabref{tab:ABJM-spin-chain-charges} according to
\be \label{eqn:string-gauge-charges}
  J_{\varphi_1} = \half p_1
  \comma
  J_{\varphi_2} = \half p_2
  \comma
  J_\psi = q + \half (p_1 + p_2)
  \; .
\ee
These relations are important for identifying classical strings with gauge theory operators. It also suggests a parametrization of $\CP^3$ inside $\Complex^4$ in terms of the embedding coordinates
\begin{align}
  y^1 & = \cos\xi \, \cos\tfrac{\theta_1}{2} \, e^{ \, i(+\varphi_1+\psi)/2 } &
  y^3 & = \sin\xi \, \cos\tfrac{\theta_2}{2} \, e^{ \, i(+\varphi_2-\psi)/2 } \\
  y^2 & = \cos\xi \, \sin\tfrac{\theta_1}{2} \, e^{ \, i(-\varphi_1+\psi)/2 } &
  y^4 & = \sin\xi \, \sin\tfrac{\theta_2}{2} \, e^{ \, i(-\varphi_2-\psi)/2 } \nn
\end{align}
which can be identified one-to-one with the scalar fields $Y^A$ of ABJM theory.

%%%%%%%%%%%%%%%%%%%%%%%%%%%%%%%%%%%%%%%%%%%%%%%%%%%%%%%%%%%%%%%%%%%%%%%%%%%%%%%%
\paragraph{Worldsheet spectrum.} In order to relate the string description to the spin-chain picture, we need to quantize the worldsheet theory. It is only known how to do this by semiclassical means, i.e.\ by expanding the string about a classical solution and quantizing the fluctuations. As can be seen from the charges, the classical string solution that corresponds to the vacuum spin-chain, or in other words to the gauge theory operator $\tr (Y^1 Y^\dagger_4)^L$ (with $L$ large so that the string becomes classical), is a point-like string that moves along the geodesic parametrized by $t = \kappa \tau$, $\psi = \kappa \tau$, located at the center of $\AdS_4$ ($\rho=0$) and the equator of $\CP^3$ ($\xi=\pi/4$), and furthermore sitting at the north pole of the first sphere ($\theta_1=0$) and at the south pole of the other sphere ($\theta_2=\pi$). Expanding the fields in fluctuations of order $\lambda^{-1/4}$ yields the mass spectrum given in \tabref{tab:fluctuations-point}.

% (Same as taking a Penrose limit of the background geometry with respect to the geodesic.)
% (If one sets longitudinal fluctuations to zero, this serves as a gauge fixing.)

The massless fluctuations $\tilde{t}$ and $\tilde{\psi}$ can be gauged away, i.e. set to zero. This is the usual light-cone gauge, $t+\psi \sim \tau$, with one light-cone direction in $\AdS_4$ and one in $\CP^3$. We are left with 4 light excitations ($\theta_{1,2}$, $\varphi_{1,2}$) from $\CP^3$ and 4 heavy excitations of which one ($\xi$) comes from $\CP^3$ and the other three ($x_{1,2,3}$) from $\AdS_4$. For the eight physical fermions the same pattern is found: 4 light excitations of mass $\kappa/2$ and 4 heavy excitations of mass $\kappa$.

These worldsheet modes transform in definite representations of the residual symmetry group $\grSU(2|2)\times\grU(1)_{\mathrm{extra}}$ that is left after fixing the light-cone gauge \cite{Bykov:2009jy}. The light fields form two $(2|2)$-dimensional supermultiplets \cite{Zarembo:2009au}
\begin{subequations} \label{eqn:worldsheet-particles}
\be
\mbox{``$A$''-particles:} && (X^a,\psi_\alpha) \; , \label{eqn:worldsheet-A-particle} \\[1mm]
\mbox{``$B$''-particles:} && (X^\dagger_a,\psi^{\dagger \alpha}) \; , \label{eqn:worldsheet-B-particle}
\ee
\end{subequations}
where $a=1,2$ and $\alpha=1,2$ are $\grSU(2)_{G}\times\grSU(2)_{r}$ indices. The doublet of complex scalars $X^a$ is a combination of $\theta_{1,2}$ and $\varphi_{1,2}$, and the fermions are written in terms of a complex spinor $\psi_\alpha$. These two supermultiplets correspond precisely to the $A$- and $B$-particles \eqref{eqn:spin-chain-particles} in the spin-chain picture, respectively!

The heavy fields form one $(1|4|3)$-dimensional supermultiplet $(\xi,\chi^a_\alpha, x_{1,2,3})$ \cite{Zarembo:2009au}. The bosonic components are literally the coordinates used above, and the fermionic component is a doublet of Majorana spinors. These heavy fields, however, do not count as independent excitations in the spin-chain description, they are rather an artifact of the above analysis which is done at infinite coupling $\lambda$. When going to finite coupling they ``dissolve'' into two light particles \cite{Zarembo:2009au}. At the technical level this is seen by looking at which particle poles appear in Green's functions at \emph{not} strictly infinite coupling \cite{Zarembo:2009au,Sundin:2009zu}. The first observation is that in the free theory the pole for the heavy particles with mass $\kappa$ coincides with the branch point of the branch cut that accounts for the pair production of two light modes with mass $\tfrac{\kappa}{2}$ each. When interactions are turned on, i.e. when $1/\sqrt{\lambda}$ corrections are considered, the pole moves into the branch cut, and the statement is that the exact propagator has a branch cut only.

%%%%%%%%%%%%%%%%%%%%%%%%%%%%%%%%%%%%%%%%%%%%%%%%%%%%%%%%%%%%%%%%%%%%%%%%%%%%%%%%
\paragraph{Giant magnons.} As we have just seen, the worldsheet fluctuations match the spin-chain excitations, but only as far as their charges are concerned. The dispersion relation of the worldsheet excitations is relativistic rather than periodic as in \eqref{eqn:general-disp-rel}. In order to see the periodic dispersion relation also on the string theory side, macroscopically many quanta must be excited. The result are classical string solutions known as giant magons \cite{Hofman:2006xt}, or dyonic giant magnons \cite{Dorey:2006dq,Chen:2006gea} if they have at least two non-zero angular momenta. The dispersion relation of all dyonic giant magnons are of the form \eqref{eqn:general-disp-rel} for appropriate values for $Q$.

The variety of giant magnons in $\CP^3$ is somewhat larger than in $\Sphere^5$. The simplest types are obtained by embedding the HM giant magnon \cite{Hofman:2006xt} into subspaces of $\CP^3$ \cite{Gaiotto:2008cg} (see also \cite{Abbott:2008qd}). There are two essentially different choices: one may either pick a proper two-sphere inside $\CP^3$ or a two-sphere with antipodes identified. According to these subspaces the former choice leads to what is called the $\CP^1$ ($\cong\Sphere^2$) giant magnon \cite{Gaiotto:2008cg} and the latter choice to the so-called $\RP^2$ ($\cong \Sphere^2/\Integers_2$) giant magnon \cite{Gaiotto:2008cg,Grignani:2008is}.

The $\RP^2$ giant magnon is in fact a threshold bound state of two HM giant magnons, one inside each of the $\Sphere^2$s parametrized by $(\theta_1,\varphi_1)$ and $(\theta_2,\varphi_2)$ in \eqref{eqn:metric-CP3} \cite{Grignani:2008is}. Therefore this kind of giant magnon is sometimes referred to as the $\Sphere^2\times\Sphere^2$ magnon or as the $\grSU(2)\times\grSU(2)$ magnon. This is, however, somewhat misleading as the two constituent magnons do not move independently.

The dyonic generalization of the $\CP^1$ giant magnon moves in a $\CP^2$ subspace of $\CP^3$ and was found for momentum $p=\pi$ in \cite{Abbott:2009um} and for general momenta in \cite{Hollowood:2009sc}. This giant magnon does not have an analogue in $\AdS_5\times\Sphere^5$. The $\CP^2$ dyonic giant magnons are in one-to-one correspondence with the elementary spin chain excitations \eqref{eqn:spin-chain-particles}: the polarizations of the giant magnons match the flavors of the excitations \cite{Hatsuda:2009pc}. In \cite{Hatsuda:2009pc} it has also been shown, that the classical phase shifts in the scattering of these dyonic giant magnons are consistent
%(up to gauge dependent terms!?)
with the S-matrix proposed by \cite{Ahn:2008aa}. The general scattering solutions of $N$ giant magnons have also been known since very recently \cite{Kalousios:2010ne}, in fact for the much wider context of giant magnons on $\CP^n$, $\grSU(n)$ and $\Sphere^n$ \cite{Hollowood:2009tw}.

The dyonic generalization of the $\RP^2$ giant magnon moves in a $\RP^3$ subspace of $\CP^3$ and was found in \cite{Ahn:2008hj}. This giant magnon is the CDO dyonic giant magnon on $\Sphere^3$ \cite{Chen:2006gea} embedded into $\RP^3$. It can be regarded as a composite of two $\CP^2$ dyonic magnons with equal momenta \cite{Hatsuda:2009pc}. Finally, by the dressing method one can also find a two-parameter one-charge solution \cite{Hollowood:2009sc,Kalousios:2009mp,Suzuki:2009sc}.

%%%%%%%%%%%%%%%%%%%%%%%%%%%%%%%%%%%%%%%%%%%%%%%%%%%%%%%%%%%%%%%%%%%%%%%%%%%%%%%%
%%%%%%%%%%%%%%%%%%%%%%%%%%%%%%%%%%%%%%%%%%%%%%%%%%%%%%%%%%%%%%%%%%%%%%%%%%%%%%%%
\section{Solving \texorpdfstring{AdS$_4$/CFT$_3$}{AdS4/CFT3} using integrability}
\label{sec:integrability}

In this section, we will briefly discuss those aspects of the methods employed to solve the AdS$_4$/CFT$_3$ model that differ from the ones in the AdS$_5$/CFT$_4$ case. For an introduction to these tools, we refer to the other chapters of this review. For the Bethe ansatz see \cite{chapABA}, for the S-matrix see \cite{chapSMat}, for the algebraic curve see \cite{chapCurve}, and for the thermodynamic Bethe ansatz and the Y-system see \cite{chapTBA,chapTrans}.

%%%%%%%%%%%%%%%%%%%%%%%%%%%%%%%%%%%%%%%%%%%%%%%%%%%%%%%%%%%%%%%%%%%%%%%%%%%%%%%%
\paragraph{Asymptotic Bethe equations.} The Bethe equations for the two-loop $\grSU(4)$ sector were derived within the algebraic Bethe ansatz scheme in \cite{Minahan:2008hf}, where also the extension of the Bethe equations to the full theory, though still at one loop, were conjectured. The form of these equations is quite canonical and the couplings between the Bethe roots is encoded in the Dynkin diagram of $\grOSp(6|4)$, see \tabref{tab:nutshell}. The all-loop extension of the Bethe equations was conjectured in \cite{Gromov:2008qe}.

The fact that we now have two types of momentum carrying roots---call them $u$ and $v$---means that the conserved charges are given by sums over all roots of both of these kinds
\be \label{eqn:BA-charges}
  Q_n = \sum_{j=1}^{K_u} q_n(u_j) + \sum_{j=1}^{K_v} q_n(v_j) \; ,
\ee
where $q_n$ is the charge carried by a single root. The spin-chain energy, or anomalous dimension, or string light-cone energy, is the second charge $E = h(\lambda) Q_2$. The other Bethe roots---call them $r$, $s$, and $w$---are auxiliary roots and influence the spectrum only indirectly through their presence in the Bethe equations. 

The $\grSU(2)\times\grSU(2)$ sector is given by only exciting the momentum carrying roots. The $\grSU(4)$ sector uses the roots $u$, $v$, $r$, though this sector is only closed at two loops. The four components of an $A$-particle, cf.\ \eqref{eqn:spin-chain-particles} and \eqref{eqn:worldsheet-particles}, correspond to the states with one $u$ root and excitation numbers $\{K_r,K_s,K_w\} = \{0,0,0\}$, or $\{1,0,0\}$, or $\{1,1,0\}$, or $\{1,1,1\}$ for the auxiliary roots. The same holds for the $B$-particle if the $u$-root is replaced by one of type $v$. This accounts for all light excitations. The heavy excitations are given by a stack of one of each kind of the momentum carrying roots. This is the Bethe ansatz way of seeing that the heavy excitations are compounds. 

This Bethe ansatz has been put to a systematic test by comparing the predicted eigenvalues to the direct diagonalization of the spin-chain Hamiltonian for various length-4 and length-6 states at two loops \cite{Papathanasiou:2009zm}.

%%%%%%%%%%%%%%%%%%%%%%%%%%%%%%%%%%%%%%%%%%%%%%%%%%%%%%%%%%%%%%%%%%%%%%%%%%%%%%%%
\paragraph{S-Matrix.} It has been shown that the proposed all-loop Bethe ansatz can be derived from an exact two-particle S-matrix \cite{Ahn:2008aa}. The alternating nature of the spin-chain, naturally breaks the S-matrix up into pieces: interactions between two $A$-particles, between two $B$-particles, and between one of each kind \cite{Ahn:2008aa}, where each piece is proportional to the old and famous $\grSU(2|2)$ S-matrix \cite{Beisert:2005tm,Arutyunov:2006yd} from AdS$_5$/CFT$_4$. Crossing symmetry relates $AA$- and $BB$- to $AB$-scattering and therefore does not fix the overall scalar factor for any of them uniquely. A solution that is consistent with the Bethe equations was made in \cite{Ahn:2008aa} and uses the BES dressing phase \cite{Beisert:2006ez}.

This S-matrix does not have poles that correspond to the heavy particles, which is in line with them not being asymptotic states. The heavy particles occur, however, as intermediate states. That is seen from the fact that they appear as internal lines in the Feynman diagrams that are used to derive the worldsheet S-matrix from scattering amplitudes \cite{Zarembo:2009au}.

The S-matrix has the peculiarity that the scattering of $A$- and $B$-particles is reflectionless \cite{Ahn:2008tv}. Though at first unexpected, this property has been confirmed perturbatively at weak \cite{Ahn:2009zg} and at strong coupling \cite{Zarembo:2009au}. This reflectionlessness would follow straightforwardly if one assumes that the two terms in \eqref{eqn:BA-charges} were individually conserved \cite{Ahn:2009tj}.

%%%%%%%%%%%%%%%%%%%%%%%%%%%%%%%%%%%%%%%%%%%%%%%%%%%%%%%%%%%%%%%%%%%%%%%%%%%%%%%%
\paragraph{Algebraic curve.} The algebraic curve for the AdS$_4$/CFT$_3$ duality was constructed from the string coset sigma-model in \cite{Gromov:2008bz}. It is a ten-sheeted Riemann surface $q(x)$ whose branches---or quasi-momenta---are pairwise related $q_{1,2,3,4,5}=-q_{10,9,8,7,6}$. The physical domain is defined for spectral parameter $\abs{x}>1$. The values of the quasi momenta within the unit circle are related to their values outside it by an inversion rule \cite{Gromov:2008bz}. Branch cut and pole conditions are identical to the ones in the AdS$_5$/CFT$_4$ case. The Virasoro constraints demand that the quasi momenta $q_1,\ldots,q_4$ all have a pole with the same residue at $x=1$ and another one at $x=-1$, while the quasi momentum $q_5$ cannot have a pole at $x=\pm1$.

For a given algebraic curve, the charges of the corresponding string solution are encoded in the large $x$ asymptotics. E.g.\ the curve
\be \label{eqn:vacuum-curve}
  q_{1}(x) = \ldots = q_{4}(x) = \frac{L}{2g} \frac{x}{x^2-1}
  \comma
  q_5(x) = 0 \; .
\ee
carries the charges $(\Delta_0,S,J_{\varphi_1},J_{\varphi_2},J_{\psi}) = (L,0,\frac{L}{2},\frac{L}{2},L)$ and $\delta\Delta = 0$ of $\tr(Y^1 Y^\dagger_4)^L$ and thus corresponds to the vacuum. String excitations are represented by additional poles that connect the various branches. A dictionary between the polarizations of the excitations and the different branch connections is given in \cite{Gromov:2008bz}. The light modes can be recognized as those which connect a non-trivial sheet with a trivial sheet in \eqref{eqn:vacuum-curve}, and the heavy modes are those which connect two non-trivial sheets.

%%%%%%%%%%%%%%%%%%%%%%%%%%%%%%%%%%%%%%%%%%%%%%%%%%%%%%%%%%%%%%%%%%%%%%%%%%%%%%%%
\paragraph{Thermodynamic Bethe ansatz and Y-system.} The Y-system for the $\grOSp(6|4)$ spin-chain was conjectured along with the corresponding equations for AdS$_5$/CFT$_4$ in \cite{Gromov:2009tv}. A derivation of the Y-system, i.e. writing down the asymptotic Bethe ansatz at finite temperature for the mirror theory, formulating the string hypothesis, and Wick rotating back to the original theory, was performed in \cite{Bombardelli:2009xz} and \cite{Gromov:2009at}, and a modification of the original conjecture was found.

%%%%%%%%%%%%%%%%%%%%%%%%%%%%%%%%%%%%%%%%%%%%%%%%%%%%%%%%%%%%%%%%%%%%%%%%%%%%%%%%
%%%%%%%%%%%%%%%%%%%%%%%%%%%%%%%%%%%%%%%%%%%%%%%%%%%%%%%%%%%%%%%%%%%%%%%%%%%%%%%%
\section*{Acknowledgements}

I am very happy to thank T.~McLoughlin and O.~Ohlsson Sax for numerous very helpful discussions. Part of this review was written while I was still affiliated with the Princeton Center for Theoretical Science whom I thank for their support.

%%%%%%%%%%%%%%%%%%%%%%%%%%%%%%%%%%%%%%%%
\phantomsection
\addcontentsline{toc}{section}{\refname}
\bibliography{chapters,TK-N6CS-IV3}
\bibliographystyle{nb}

\end{document}